\documentclass[aps,prl,preprint]{revtex4}

\usepackage{epsfig}

\draft

\begin{document}
\author{Hong-shi Zong$^{1,2}$, Xiao-hua Wu$^{3}$, Feng-yao Hou$^{1}$,
En-guang Zhao$^{2,4}$}
\address {$^{1}$ Department of Physics, Nanjing University, Nanjing 210093, P. R. China} 
\address{$^{2}$ CCAST(World Laboratory), P.O. Box 8730, Beijing 100080, P. R. China}\address{$^{3}$  Department of Physics, Sichuan University,
Chengdu 610064, P. R. China}
\address{$^{4}$ Institute of Theoretical Physics, Academia Sinica, P.O. Box 2735, Beijing 100080, P. R. China}

\title{The Explicit and Dynamical Chiral Symmetry Breaking in an Effective Quark-quark Interaction Model}

\begin{abstract}
A method for obtaining the small current quark mass effect on the dressed quark 
propagator from an effective quark-quark interaction model is developed. Within this approach
both explicit and dynamical chiral symmetry breaking has been analyzed.
A comparison with previous results is given.

\bigskip

Key-words:  GCM, Explicit and dynamical chiral symmetry breaking.

\bigskip

E-mail: zonghs@chenwang.nju.edu.cn.

\bigskip

PACS Numbers: 24.85.+p, 13.60.Le

\end{abstract}

\maketitle

The dynamical chiral symmetry breaking in QCD is a typical nonperturbative phenomenon, where the standard perturbative scheme is not valid. While lattice QCD is the most straightforward and solid approach among nonperturbative studies of QCD, analytical studies with effective theories are useful alternatives and often shed more light on nonperturbative phenomena than lattice QCD. The calculation of the last twenty years show that the global color symmetry model(GCM) provides a successful description of various nonperturbative aspect of strong interaction physics[1-4]. We naturally expect that it might be a useful model in the study of the small current quark mass effects on the dressed quark propagator which manifests the behavior of the explicit and dynamical chiral symmetry breaking.

The previous study of dressed quark propagator for the light quarks $u$ and $d$ has been done often in the chiral limit, where the current quark mass is set to zero. However, in a more realistic situation, i.e., quarks have small but finite masses representing explicit symmetry breaking which play an important role in the study of QCD phase structure.
 It is interesting to address small current quark mass effects on the dressed quark propagator further. Up to this let us start from the Euclidean action of the GCM at small but finite current quark mass $m$:
\begin{eqnarray}
&&S_{GCM}[\bar{q},q;m]\nonumber\\
&&=\int d^{4} x \left\{\bar{q}(x)[\gamma\cdot\partial_{x}+m]q(x)\right\}+\int d^{4}x d^{4} y\left[\frac{g^2_{s}}{2} j^{a}_{\mu}(x) D^{ab}_{\mu\nu}(x-y)j^{b}_{\nu}(y)\right],
\end{eqnarray}
where $j^{a}_{\mu}(x)=\bar{q}(x)\gamma_{\mu}\frac{\lambda^{a}_{c}}{2}q(x)$  denotes the color octet vector current and $g^2_{s}D^{ab}_{\mu\nu}(x-y)$ is the dressed model gluon propagator in GCM. Without loss of generality, we will employ the Landau gauge $D^{ab}_{\mu\nu}(q)=\delta_{ab}\left(\delta_{\mu\nu}-\frac{q_{\mu}q_{\nu}}{q^2}\right)D(q^2)$ for the model gluon propagator.

Introducing an auxiliary bilocal field $B^{\theta}(x,y)$ and applying the standard bosonization procedure the partition function of GCM[1,2]
\begin{equation}
{\cal{Z}}[m]=\int{\cal{D}} \bar{q}{\cal{D}}q e^{-S_{GCM}[\bar{q},q;m]}
\end{equation}
can be rewritten in terms of the bilocal fields $B^{\theta}(x,y)$
\begin{equation}
{\cal{Z}}[m]=\int{\cal{D}} B^{\theta} e^{-S_{eff}[B^{\theta};m]}
\end{equation}
with the effective bosonic action
\begin{equation}
S_{eff}[B^{\theta};m]=-Tr ln {\cal{G}}^{-1}[B^{\theta};m]+I[B^{\theta}]
\end{equation}
and the quark operator
\begin{equation}
{\cal{G}}^{-1}[B^{\theta};m]=[\gamma\cdot\partial_{x}+m]\delta(x-y)
+\Sigma(x,y),
\end{equation}
where the explicit expressions of $I[B^{\theta}]$ and $\Sigma(x,y)\equiv\Lambda^{\theta} B^{\theta}(x,y)$ can be found in Ref.[5]. The matrices $\Lambda^{\theta}=D^{a}\otimes C^{b}\otimes F^{c}$ is determined by Fierz transformation in Dirac, color and flavor space of the current current interaction in Eq.(1).

In the mean-field approximation, the fields $B^{\theta}(x,y)$ are substituted simply by their vacuum value $B^{\theta}_{0}(x,y)$, which is defined as
\begin{equation}
\frac{\delta{S}_{eff}}{\delta{B}}{\mid}_{B_{0}}=0.
\end{equation}
Employing the stationary condition Eq.(6), and reversing the Fierz transformation, we have
quark self-energy $\Sigma_{0}(x,y)$ at the mean-field approximation
\begin{equation}
\Sigma_{0}[m](x,y)=\frac{4}{3}g^2_{s} D_{\mu\nu}(x,y)\gamma_{\mu}
{\cal{G}}_{0}[m](x,y)\gamma_{\nu}.
\end{equation}
It should be noted that both $B^{\theta}_{0}(x,y)$ and ${\cal{G}}^{-1}_{0}(x,y)$ are dependent on the small current quark mass $m$. If $m$ is switched off, ${\cal{G}}_{0}[m]$ goes into the dressed vacuum quark propagator in the chiral limit $G\equiv{\cal{G}}_{0}[m=0]$, which has the 
decomposition
\begin{equation}
G^{-1}(p)=i\gamma\cdot p+\Sigma_{0}(p)=i\gamma\cdot p A(p^2)+B(p^2)
\end{equation}
with
\begin{equation}
\Sigma_{0}(p)=\int d^4 x e^{i p\cdot x}[\Lambda^{\theta} B^{\theta}_{0}(x)]=\frac{4}{3}\int \frac{d^4 q}{(2\pi)^4} g^2_{s}D_{\mu\nu}(p-q)\gamma_{\mu}G(q)\gamma_{\nu}
\end{equation}
where the self energy functions $A(p^2)$ and $B(p^2)$ are
determined by the rainbow Dyson-Schwinger equation(DSE) in the chiral limit:
\[
[A(p^2)-1]p^2=\frac{4}{3}\int \frac{d^{4}q}{(2\pi)^4}g^2_{s} D(p-q)
\frac{A(q^2)}{q^2A^2(q^2)+B^2(q^2)}\left[p\cdot q+2\frac{p\cdot(p-q)~q\cdot(p-q)}{(p-q)^2}\right],
\]
\begin{equation}
B(p^2)=4\int \frac{d^{4}q}{(2\pi)^4}g^2_{s} D(p-q)
\frac{B(q^2)}{q^2A^2(q^2)+B^2(q^2)}.
\end{equation}

Here we should stress that the $B(p^2)$ in Eq.(10) has two
qualitatively distinct solutions. The ``Nambu-Goldstone''
solution, for which $B(p^2)\neq 0$,
describes a phase, in which: 1) chiral symmetry is dynamically
broken, because one has a nonzero quark mass function; and 2) the
dressed quarks are confined, because the propagator described by
these functions does not have a Lehmann representation. The
other solution, the ``Wigner'' one, $B(p^2)\equiv 0$,
describes a phase, in which chiral symmetry is not broken. In
``Wigner'' phase, the Dyson-Schwinger equation(10) reduces to:
\begin{equation}
[A'(p^2)-1]p^2=\frac{4}{3}\int \frac{d^{4}q}{(2\pi)^4}g_{s}^2 D(p-q)
\left[p\cdot q+2\frac{p\cdot(p-q)~q\cdot(p-q)}{(p-q)^2}\right]
\frac{1}{q^2A'(q^2)},
\end{equation}
where $A'(p^2)$ denotes the dressed quark vector self energy function in the ``Wigner'' phase. Therefore, in chiral limit, the dressed quark propagator in the ``Wigner'' phase can be
written as $G^{(W)}(q)=\frac{-i\gamma\cdot q}{A'(q^2)q^2}$.

Let us now study the small current quark mass dependence of the dressed quark propagator. To this, for the light quarks $u$ and $d$ one can expands ${\cal{G}}^{-1}_{0}[m]$ in powers of $m$ as follows
\begin{equation}
{\cal{G}}^{-1}_{0}[m]=\left.{\cal{G}}^{-1}_{0}[m]\right|_{m=0}+\left. \frac{\delta {\cal{G}}^{-1}_{0}[m]}{\delta {m}} \right|_{m=0}m+
{\cal{O}}(m^2)=G^{-1}+m\Gamma+{\cal{O}}(m^2),
\end{equation}
which leads to the formal expansion
\begin{equation}
{\cal{G}}_{0}[m]=G-m G\Gamma G+\cdot\cdot\cdot,
\end{equation}
with $\Gamma$
\begin{equation}
\Gamma(y_1,y_2)=\left. \frac{\delta {\cal{G}}^{-1}_{0}[m](y_1,y_2)}
{\delta m} \right|_{m=0}.
\end{equation}
In coordinate space the dressed vertex $\Gamma(x,y)$ is given as the 
derivative of the inverse quark propagator ${\cal{G}}^{-1}_{0}[m]$ with respect to the $m$. 

Taking the derivative in Eq.(5) and putting it into Eq.(12), we have
\begin{equation}
\Gamma(y_1,y_2)=\delta(y_1-y_2)+
\left. \frac{\delta\Sigma_{0}[m](y_1,y_2)}{\delta{m}} \right|_{m=0}.
\end{equation}

Substituting Eq.(7) and (13) into Eq.(15), we have the inhomogeneous ladder Bethe-Salpeter equation(BSE) of scalar vertex $\Gamma$, which reads
\begin{eqnarray}
\Gamma(y_1,y_2)&&=\delta(y_1-y_2)\nonumber\\
&&-\frac{4}{3}g^2_{s}D_{\mu\nu}(y_1-y_2)\int du_{1}du_{2}\gamma_{\mu}G(y_1,u_1)
\Gamma(u_1,u_2)G(u_2,y_2)\gamma_{\nu}.
\end{eqnarray}

Fourier transform of Eq.(16) leads then to the momentum space form of $\Gamma$
\begin{equation}
\Gamma(p,0)
=1-\frac{4}{3}\int\frac{d^4 q}{(2\pi)^4}g^2_{s}D_{\mu\nu}(p-q)\gamma_{\mu}G(q)\Gamma
(q,0)G(q)\gamma_{\nu}.
\end{equation}
As was shown above, both the rainbow DSE(10) and the ladder BSE(17) can be consistently derived from the action of the GCM(1) at the mean field level.

From Lorentz structure, the most general form for the $\Gamma$ which fulfills Eq.(17), 
reads:
\begin{equation}
\Gamma(p,0)=i\gamma\cdot p~C(p^2)+D(p^2).
\end{equation}
Substituting Eq.(18) into Eq.(17), we have the coupled integral equation for $C(p^2)$ and $D(p^2)$:
\begin{eqnarray}
P^2C(p^2)&&=-\frac{4}{3}\int\frac{d^4q}{(2\pi)^4} g^2_{s}D(p-q)\left[p\cdot q+\frac{2p\cdot(p-q)~q\cdot(p-q)}{(p-q)^2}\right]\times\nonumber\\
&&\left\{\frac{\left[2A(q^2)B(q^2)D(q^2)+q^2A^2(q^2)C(q^2)-C(q^2)B^2(q^2)\right]}{\left[q^2A^2(q^2)+B^2(q^2)\right]^2}\right\},\nonumber\\
D(p^2)&&=1-4\int\frac{d^4q}{(2\pi)^4} g^2_{s}D(p-q)\times\nonumber\\
&&\left\{\frac{\left[2q^2A(q^2)B(q^2)C(q^2)-q^2A^2(q^2)D(q^2)+D(q^2)B^2(q^2)\right]}{\left[q^2A^2(q^2)+B^2(q^2)\right]^2}\right\}.
\end{eqnarray}

In ``Wigner'' phase, $B(p^2)\equiv 0$. Substituting $B(p^2)=0$ into Eq.(19), we have the decoupled integral equation for $C'(p^2)$ and $D'(p^2)$ in ``Wigner'' phase:
\begin{eqnarray}
P^2C'(p^2)&&=-\frac{4}{3}\int\frac{d^4q}{(2\pi)^4} g^2_{s}D(p-q)\left[p\cdot q+\frac{2p\cdot(p-q)~q\cdot(p-q)}{(p-q)^2}\right]\frac{C'(q^2)}{q^2A'^2(q^2)},\nonumber\\
D'(p^2)&&=1+4\int\frac{d^4q}{(2\pi)^4} g^2_{s}D(p-q)
\frac{D'(q^2)}{q^2A'^2(q^2)}.
\end{eqnarray}
From Eq.(20), it is easy to find $C'(p^2)\equiv 0$ in the ``Wigner'' phase.

Based on the above coupled integral equation(19) and Eq.(10), by means of numerical studies, we can get the nonperturbative scalar vertex $\Gamma$, which is useful for the studies of the small current quark mass effects on the dressed quark propagator. It should be noted that the above approach for getting the nonperturbative scalar vertex $\Gamma$ has been proven to be very useful for the studies of nonperturbative vector, axial vector vertex[6-8] and the chemical potential dependence of the dressed quark propagator[9]. So far, at the mean field level, we have completed the derivation of the first order dependence of ${\cal{G}}^{-1}_{0}[m]$ upon $m$ in ``Nambu-Goldstone'' and ``Wigner'' phase separately;
\begin{eqnarray}
{{\cal{G}}}_{0}^{(NG)^{-1}}[m]&\equiv& {{\cal{G}}}_{0}^{(NG)^{-1}}(D)+
{{\cal{G}}}_{0}^{(NG)^{-1}}(E)+{\cal{O}}(m^2)\nonumber\\
&=&i\gamma\cdot p~A(p^2)+B(p^2)+m\left[i\gamma\cdot p~C(p^2)+D(p^2)\right]
+{\cal{O}}(m^2),
\end{eqnarray}
\begin{eqnarray}
{{\cal{G}}}_{0}^{(W)^{-1}}[m]&\equiv& {{\cal{G}}}_{0}^{(W)^{-1}}(D)+
{{\cal{G}}}_{0}^{(W)^{-1}}(E)+{\cal{O}}(m^2)\nonumber\\
&=&i\gamma\cdot p~A'(p^2)+m D'(p^2)+{\cal{O}}(m^2).
\end{eqnarray}
Here ${{\cal{G}}}_{0}^{(NG)^{-1}}(D)\equiv i\gamma\cdot p~A(p^2)+B(p^2)$ is the contribution of dynamical symmetry breaking and is independent of the current quark mass $m$. 
${{\cal{G}}}_{0}^{(NG)^{-1}}(E)\equiv m\left[i\gamma\cdot p~C(p^2)+D(p^2)\right]$ is the contribution of the explicit symmetry breaking and vanishes if the current quark mass $m=0$. It is now apparent that the above approach for getting the nonperturbative scalar vertex has the advantage that the effects of explicit and dynamical chiral breaking can be analyzed separately.

As is shown in Eq.(21), for $m\not= 0$ the dressed-quark self energies in general acquire not only the scalar part $D(p^2)$ but also the vector part $C(p^2)$ driven by the current quark mass $m$. One often miss the contribution of the vector part $C(p^2)$ in the previous study of dressed quark propagator for light quarks.

In order to have a qualitative understanding of the contribution of $C(p^2)$, $D(p^2)$, and  $D'(p^2)$, a particularly simple and useful model of the dressed gluon two-point 
function[10] is used as follows:
\begin{equation}
g_{s}^2 D_{\mu\nu}(p-q)=4\pi^4\eta^2\left[\delta_{\mu\nu}-\frac{(p-q)_{\mu}(p-q)_{\nu}}{(p-q)^2}\right]\delta^{(4)}(p-q).
\end{equation}
Here the scale parameter $\eta$ is a measure of the strength of the infrared slavery effect.
It should be noted that the model gluon propagator(23) is an infrared-dominant model that does not represent well the behavior of $g_{s}^2D_{\mu\nu}(p)$ away from $p^2\simeq 0$.

Then substituting Eq.(23) into Eq.(10) and (11), we have the ``Nambu-Goldstone'' solution; 
\begin{eqnarray}
B(p^2)&=&(\eta^2-4 p^2)^{\frac{1}{2}},~~~~~~~A(p^2)=2 ~~~~for~~~~ 
p^2 < \frac{\eta^2}{4},\nonumber\\
B(p^2)&=&0,~~~~A(p^2)=\frac{1}{2}\left[1+(1+\frac{2\eta^2}{p^2})^{\frac{1}{2}}
\right]~~~for~~~~ p^2\geq\frac{\eta^2}{4},
\end{eqnarray}
and the ``Wigner'' solution;
\begin{eqnarray}
B'(p^2)&=&0,~~~~A'(p^2)=\frac{1}{2}\left[1+(1+\frac{2\eta^2}{p^2})^{\frac{1}{2}}
\right].
\end{eqnarray}

With the model of the dressed gluon propagator specified in Eq.(23) and Eqs.(24,25), from Eq.(19) and (20) we have the following simple form of the $C(p^2)$, $D(p^2)$ 
\begin{eqnarray}
D(p^2)&=&\frac{p^2+\eta^2+p^2(1+2\eta^2/p^2)^{\frac{1}{2}}}{p^2-\eta^2+p^2(1+2\eta^2/p^2)^{\frac{1}{2}}},~~~~~C(p^2)=0~~~p^2\geq\frac{\eta^2}{4},\\
D(p^2)&=&\frac{1}{2}\frac{\eta^2+8p^2}{\eta^2-4p^2},~~~~C(p^2)=-\frac{2(\eta^2-4p^2)^{\frac{1}{2}}}{\eta^2-4p^2},~~~~P^2 < \frac{\eta^2}{4},\nonumber
\end{eqnarray}
and $C'(p^2)$, $D'(p^2)$
\begin{eqnarray}
D'(p^2)&=&\frac{p^2+\eta^2+p^2(1+2\eta^2/p^2)^{\frac{1}{2}}}{p^2-\eta^2+p^2(1+2\eta^2/p^2)^{\frac{1}{2}}},~~~~~C'(p^2)\equiv 0.
\end{eqnarray}

At large momentum region, i.e., $p^2\rightarrow\infty$, from Eq.(26), we have 
\begin{eqnarray}
D(p^2)\rightarrow 1, ~~~~~~~~~C(p^2)=0.
\end{eqnarray}
This asymptotic behavior is quite different from that given by Ref.[11]. This is because the model gluon propagator specified by Eq.(23) is an infrared-dominant model that does not represent well the real large momentum behavior of $g_{s}^2D_{\mu\nu}(p)$ in QCD. This question will be further discussed by means of a more sophisticated ansatze for the gluon propagator elsewhere.

With these two ``phase'' characterized by qualitatively different momentum-dependent quark propagator(21,22), the GCM can be used to calculate the vacuum condensates and susceptibilities at the mean field level[12]. In particular we obtain the chiral quark condensate with the small current quark mass $m$:
\begin{eqnarray}
&&\langle\tilde{0}|:\bar{q}q:|\tilde{0}\rangle_{m}=-tr_{DC}
\left\{{\cal{G}}^{(NG)}_{0}[m]-{\cal{G}}^{(W)}_{0}[m]\right\}\\
&&=-\frac{3}{4\pi^2}\int^{\eta^2/4}_{0}ds s\left\{\frac{2(\eta^2-4s)[2(\eta^2-4s)^{3/2}+m(\eta^2+8s)]}{16s[(\eta^2-4s)-m(\eta^2-4s)^{1/2}]^2+[2(\eta^2-4s)^{3/2}+m(\eta^2+8s)]^2}
\right.\nonumber\\
&&~~~~~~~~~~~~~~~~~~~~~~~\left.-\frac{2m[s-\eta^2+s(1+2\eta^2/s)^{1/2}]}
{[s-\eta^2+s(1+2\eta^2/s)^{1/2}]^2+2m^2[s+\eta^2+s(1+2\eta^2/s)^{1/2}]}\right\}.\nonumber
\end{eqnarray}

Substituting $m=0$ into Eq.(29), we have the chiral quark condensate in chiral limit:
\begin{equation}
\langle\tilde{0}|:\bar{q}q:|\tilde{0}\rangle_{m=0}=-tr_{DC}
\left\{{\cal{G}}^{(NG)}_{0}[m=0]\right\}=-\frac{\eta^3}{80\pi^2}
\end{equation}

The calculated ratio $\langle\tilde{0}|:\bar{q}q:|\tilde{0}\rangle_{m}/\langle\tilde{0}|:\bar{q}q:|\tilde{0}\rangle_{m=0}$ can be seen from Fiq.1. In Fig.1, we see that $\langle\tilde{0}|:\bar{q}q:|\tilde{0}\rangle_{m}/\langle\tilde{0}|:\bar{q}q:|\tilde{0}\rangle_{m=0}$ increases with increasing $m$ within the certain range of the small current quark mass(it should be noted that our numerical result is only valid for low values of $m$). In order to show 
the contribution of the subtraction term $tr_{DC}\left\{{\cal{G}}^{(W)}_{0}[m]\right\}$, we plot the ratio $tr_{DC}\left\{{\cal{G}}^{(W)}_{0}[m]\right\}/\langle\tilde{0}|:\bar{q}q:|\tilde{0}\rangle_{m=0}$ versus $m$ in Fig.2. It is easy to find that the contribution of subtraction term can not be neglected in calculating the chiral quark condensate with finite current quark mass $m$.

\begin{center}

\epsfig{file=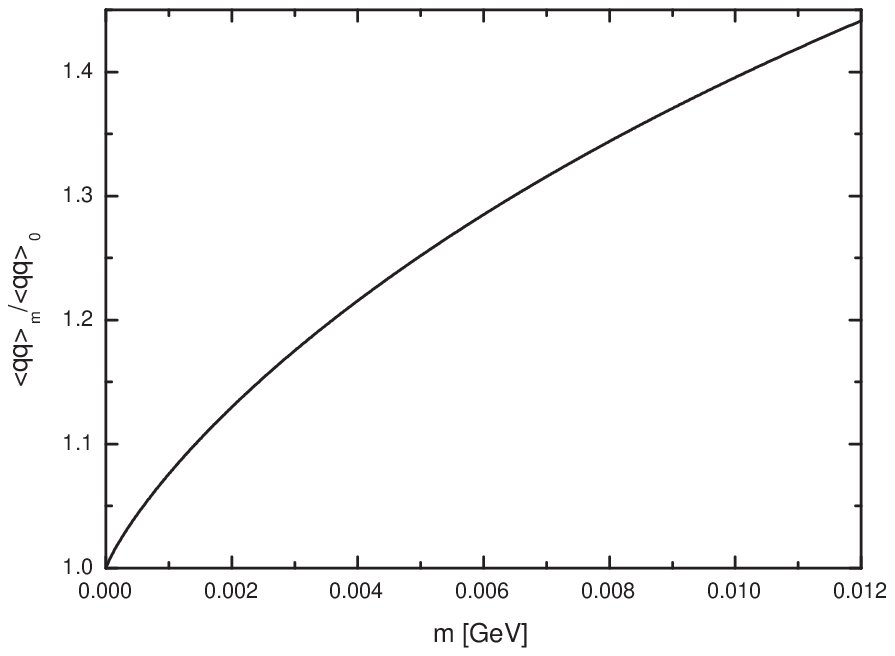, width=13cm}

\vspace{-0.7cm}

Fig.1. The ratio
$\langle\tilde{0}|:\bar{q}q:|\tilde{0}
\rangle_{m}/\langle\tilde{0}|:\bar{q}q:|\tilde{0}\rangle_{m=0}$ as a function of $m$ for $\eta^2=1.14~GeV^2$.

\epsfig{file=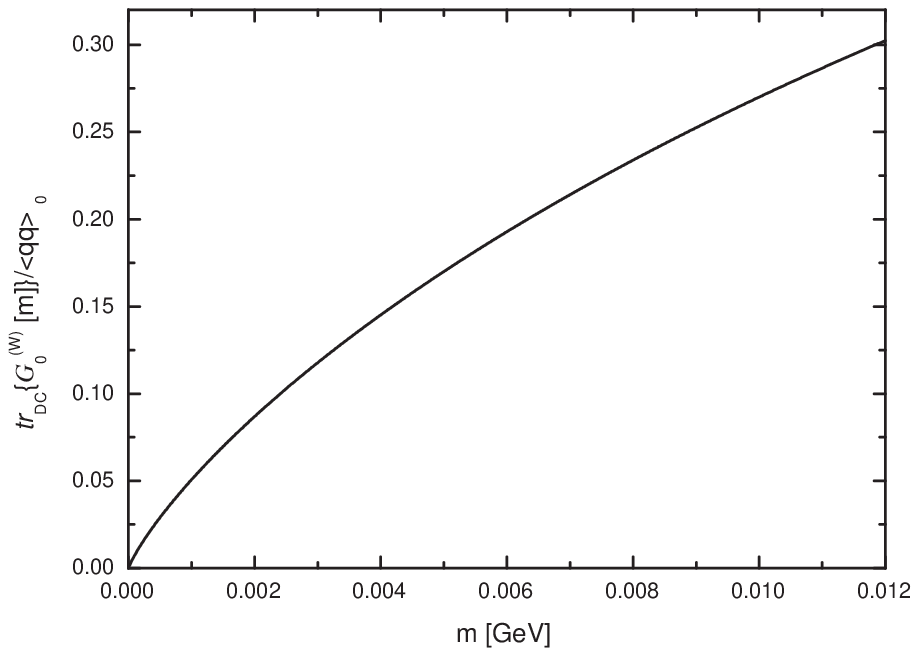, width=13cm}

\vspace{-0.7cm}

Fig.2. The ratio
$tr_{DC}\left\{{\cal{G}}^{(W)}_{0}[m]\right\}/\langle\tilde{0}|:\bar{q}q:|\tilde{0}\rangle_{m=0}$ as a function of $m$ for $\eta^2=1.14~GeV^2$.

\end{center}

To summarize: in order to discuss explicit and dynamical chiral symmetry breaking, we first provide a general recipe to calculate the small current quark mass effect on the dressed quark propagator at the mean field level in the framework of GCM. This employs a consistent treatment of dressed quark propagator $G$ and the dressed scalar vertex $\Gamma$, which are both determined from the effective quark-quark interaction by the rainbow DSE for $G$ and the inhomogeneous ladder BSE for $\Gamma$. This approach has the advantage that we can analyze the effects of explicit and dynamical chiral symmetry breaking separately. Then we use a 
simple, confining model(23) to calculate the chiral quark condenste in the case of the small current quark mass. It is found that the contribution of the subtraction term $tr_{DC}\left\{{\cal{G}}^{(W)}_{0}[m]\right\}$ can not be dropped.

\vspace*{0.8 cm}
\noindent{\large \bf Acknowledgments}

This work was supported in part by the National Natural Science Foundation
of China under Grant Nos 19975062, 10175033 and 10135030.

\vspace*{0.8 cm}
\noindent{\large \bf References}

\begin{description}
\item{[1]} R. T. Cahill and C. D. Roberts, Phys. Rev. {\bf D32}, 2419 (1985).
\item{[2]} P. C. Tandy, Prog. Part. Nucl. Phys. 39, 117 (1997); R. T. Cahill and S. M. Gunner, Fiz. {\bf B7}, 17 (1998), and references therein.
\item{[3]} C. D. Roberts and A. G. Williams, Prog. Part. Nucl. Phys. {\bf 33}, 477 (1994), and references therein.
\item{[4]} C. D. Roberts and S. M. Schmidt, Prog. Part. Nucl. Phys. {\bf 45S1}, 
1 (2000), and references therein.
\item{[5]} C. D. Roberts and R. T. Cahill, Aust. J. Phys. {\bf 40}, 499(1987).
\item{[6]} M. R. Frank, Phys. Rev. {\bf C51}, 987 (1995).
\item{[7]} T. Meissner and L. S. Kisslinger,  Phys. Rev. {\bf C59}, 986 (1999).
\item{[8]} Hong-shi Zong, Xiang-song Chen, Fan Wang, Chao-hsi Chang and En-guang Zhao, 
Phys. Rev. {\bf C66}, 015201 (2002).
\item{[9]} Hong-shi Zong, Jia-lun Ping, Wei-min Sun, Chao-hsi Chang and Fan Wang, 
nucl-th/0209015.
\item{[10]} H. J. Munczek and A. M. Nemirovsky, Phys. Rev. {\bf D28}, 181 (1983).
\item{[11]} H. Pagels, Phys. Rev. {\bf D19}, 3080 (1979).
\item{[12]} Hong-shi Zong, Jia-lun Ping, Hong-ting Yang, Xiao-fu Lu and Fan Wang, 
nucl-th/0201001.

\end{description}

\end{document}